\documentclass[a4paper,epj]{svjour}

\usepackage{graphicx}
\usepackage{bm}
\usepackage{dcolumn}
\usepackage{amssymb}

\begin{document}

\title{
  The high-pressure \boldmath$\alpha / \beta$ phase transition in lead
  sulphide (PbS)}
\subtitle{X-ray powder diffraction and quantum mechanical
  calculations}


\author{
  K. Knorr\inst{1}
  \and
  L. Ehm\inst{1}
  \and
  M. Hytha\inst{1}
  \thanks{\emph{Also at:} Institute of Physics of the Czech 
    Academy of Sciences,
    Cukrovarnicka 10, 16253 Praha 6, Czech Republic}
  \thanks{\emph{Present address:} University of Cambridge, Cavendish
    Laboratory (TCM), Madingley Road, Cambridge CB3 0HE, U.K.
    }
  \and
  B. Winkler\inst{1}
  \thanks{\emph{Present address:} Johann-Wolfgang Goethe Universit\"{a}t,
    Mineralogisches Institut, Kristallographie, Senckenberganlage 30, D
    60054 Frankfurt a.M., Germany}
  \and
  W. Depmeier\inst{1}
  }

\offprints{K. Knorr, {\tt knorr@min.uni-kiel.de}}

\institute{Christian-Albrechts-Universit\"at zu Kiel, Institut f\"ur
  Geowissenschaften, Mineralogie/Kristallographie,\\ Olshausenstr. 40,
  D-24098 Kiel, Germany
  }

\date{\today}

\abstract{ \advance\baselineskip by 6pt The high-pressure behaviour of
  PbS was investigated by angular dispersive X-ray powder diffraction
  up to pressures of 6.8 GPa. Experiments were accompanied by first
  principles calculations at the density functional theory level. By
  combining both methods reliable data for the elastic properties of
  rock-salt type $\alpha$- and high-pressure $\beta$-PbS could be
  obtained.  $\beta$-PbS could be determined to crystallise in the
  CrB-type (B33), with space group \textit{Cmcm}.  The reversible
  ferro-elastic $\alpha/\beta$ transition is of first order. It is
  accompanied by a large volume discontinuity of about 5\% and a
  coexistence region of the two phases. A gliding mechanism of \{001\}
  bilayers along one of the cubic $\langle 110 \rangle$ directions
  governs the phase transition which can be described in terms of
  group/subgroup relationships via a common subgroup, despite its
  reconstructive character. The quadrupling of the primitive unit cell
  indicates a wave vector $(0,0,\pi/a)$ on the $\Delta$-line of the
  Brillouin zone.  \PACS{ {91.60.Gf}{High-pressure behaviour} \and
    {61.50.Ks}{Crystallographic aspects of phase transformations;
      pressure effects} \and {77.84.Bw}{Elements, oxides,
      nitrides, borides, carbides, chalcogenides} \and {62.20.Dc}{Elasticity, elastic constants}} }

\maketitle


\sloppy

\section{Introduction}

Among the IV-VI narrow gap semiconductors lead sulphide, PbS, has been
a subject of interest during the last three decades, owing to its
potential technological relevance, e.g. in long-wavelength imaging, as
diode lasers \cite{Philipps2000} or thermovoltaic energy converters.
Many of the experimental and theoretical investigations have been
focused on the electronic structure \cite{Santoni92,Wei97}. Currently,
PbS attracts new interest because of its importance for the
understanding of the complex misfit layer compounds
(MX)$_{1+x}$(TX$_2$)$_m$ (M=Sn, Pb, Sb, Bi, rare earth; T=Ti, V, Cr,
Nb, Ta; X=S, Se; 0.08$<$x$<$0.28; m=1-3) \cite{Rouxel95}, of which PbS
can be regarded as a component structure. These composite structures
are single-crystal multi-layer systems, exhibiting strong
anisotropic material properties.

At ambient conditions $\alpha$-PbS crystallises in the rock salt
structure type (B1) \cite{Wasserstein51} with the lattice parameter
$a$=5.9240(4)~\AA\/ and space group $Fm\bar 3m$.  The elastic
constants of $\alpha$-PbS were determined by ultrasonic velocity
measurements \cite{Padaki81}.  Resistivity measurements under high
pressure on several lead chalcogenides revealed the existence of a
phase transition \cite{Samara62} which occurs  in PbS at 2.2 GPa.
First structural studies under high pressure were performed using X-ray
film methods \cite{Wakabayashi68}.  Later, the compression behaviour
of PbS was studied by means of high-pressure energy dispersive
X-ray powder diffraction (HP-EDX) up to 35 GPa
\cite{Peresada76,Chattopadhyay84,Chattopadhyay86,Quadri96,Jiang00}.

The phase transition at 2.2 GPa \cite{Samara62} 
to a $\beta$-phase having orthorhombic symmetry was characterised as a
first order transition
\cite{Wakabayashi68,Chattopadhyay84,Chattopadhyay86}. However, the structure
of this phase is still a matter of debate.
Two different structure
types were proposed, viz. the GeS structure type (B16) with the space
group \textit{Pbnm}
\cite{Samara62,Wakabayashi68,Quadri96,Jiang00} and the closely related,
but higher symmetric CrB structure type (B33) with the symmetry \textit{Cmcm}
\cite{Chattopadhyay84,Chattopadhyay86}. 
A definitive decision between these two possibilities has not been
possible until now
on the basis of the  experimental data available.

A second phase transition at about 25 GPa was detected by HP-EDX and
assigned to the transformation into  cubic $\gamma$-PbS with the
CsCl (B2) type structure \cite{Chattopadhyay84,Chattopadhyay86}.

Since the equation-of-state parameters for $\alpha$-PbS, the symmetry of the
$\beta$ phase and its compression behaviour are hitherto unknown
it was found worth reexamining  the PbS system under pressure.
To elucidate the structure of the intermediate $\beta$ phase, angle
dispersive X-ray powder diffraction and density functional theory
(DFT) calculations were performed. Here the results of the study of
the crystal structure of PbS under high pressure are presented.  The
mechanism of the phase transition is discussed using the
symmetry relations between the space groups of the two phases.

\section{Experimental methods}

Crystals were grown by chemical vapour transport (CVT) in evacuated
quartz tubes in a gradient furnace (1053 to 923~K) with
stoichiometrically mixed elements as starting material and iodine as
transport agent.  In the low-temperature zone cube shaped crystals
were obtained after several days with edge lengths up to 3~mm.  The
chemical composition of the crystals was checked by electron
micro-probe analysis and the phase purity was studied by conventional
X-ray powder diffraction.

The high-pressure experiments were performed using diamond anvil cells
\cite{Merrill74}. NaCl was used as internal standard for the pressure
determination by the Decker equation-of-state  \cite{Decker71}.
Several runs were performed using a 4:1
me\-tha\-nol/ethanol mixture for maintaining hydrostatic conditions. The
highest pressure obtained was 6.8~GPa. Angular dispersive diffraction
data were collected with Mo-$K\alpha$ radiation on a {\sc Mar2000} image plate
diffractometer.  The geometry correction for the radial integration of
the two-dimensional data and the transformation into standard
one-dimensional powder patterns were performed using {\sc Fit2d}
\cite{Hammersley96}.  Whole powder pattern fitting employing {\sc
  Fullprof} \cite{Rodriguez93} was performed for the precise
determination of the unit cell parameters.

\section{Computational aspects}

Geometry optimisations were performed for $\alpha$-PbS and for both
proposed structure types of $\beta$-PbS. The {\it ab initio}
plane-wave CASTEP code \cite{Payne92,Milman2000} was used which is
based on the density functional theory (DFT). The ex\-change cor\-re\-la\-tion
interaction was accounted for by using the Perdew, Burke and Ernzerhof
version of the generalised gradient approximation (PBE-GGA) of the
exchange-correlation functional \cite{Perdew92,White94}.  Ultrasoft
pseudopotentials \cite{Vanderbilt90,Kresse94} with a maximum cutoff
energy of the plane waves of 290 eV were used to describe the
electron-ion interaction.  The integration over the Brillouin zone
was performed employing the Monkhorst-Pack \cite{Monkhorst76}
sampling scheme of reciprocal space, with a distance of 0.04
\AA$^{-1}$ between the sampling points.  All structural parameters not
constrained by the space group symmetry were relaxed.

The  calculation of the elastic constants  is based on
applying  small strains to the already relaxed structure, followed by
 relaxation of the atomic positions. In order to avoid anharmonic effects the maximum
applied distortion was about 0.2 \%. Several amplitudes for
each strain component were used. Elastic constants were determined
from a linear fit of stress to the applied strain.

\section{Results and discussion}

\begin{figure}
\centering
\includegraphics[bb=115 195 445 555,angle=-90, width=.60\columnwidth]{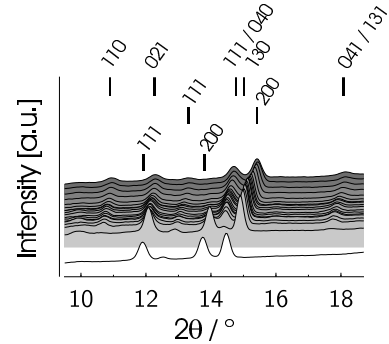}
\caption{
  Pressure dependence of diffraction patterns of PbS from room
  pressure to 6.87 GPa. The patterns are shifted with increasing
  pressure. Miller indices are given for $\alpha$-PbS (lower row),
  NaCl (middle) and, $\beta$-PbS (top row).}
\label{fig:wf}
\end{figure}

Some integrated powder patterns are presented in figure \ref{fig:wf}.
The onset of a phase transition was detected at 2.5 GPa by the
appearance of additional reflections. The new phase could be indexed
with an orthorhombic unit cell and cell parameters $a$=3.98(2)~\AA,
$b$=11.11(7)~\AA\/ and $c$=4.16(4)~\AA.  Coexistence of the two phases
was observed up to a pressure of 6 GPa.

The powder patterns obtained above 2.5 GPa could be indexed with the
same unit cell for both, B16 and B33, structure types.  A
structure refinement of the $\beta$-phase in either case turned out to
be impossible because of the
strong peak overlap with reflections stemming from the gasket
material (Inconel steel). Thus, a discrimination between both possible
structure types was not possible by the diffraction method.

\subsection{Lattice parameters and elastic properties of $\bm \alpha$- and $\bm \beta$-PbS}

\begin{table}[bt]
\caption{Parameters of Birch Murnaghan equations-of-state
  \cite{Birch78} 
  derived from experimental and calculated data and values of the linear
  compressibility along the directions of the lattice
  vectors for $\alpha$- and $\beta$-PbS. The linear compressibilities
  from the DFT calculations were derived from the elastic compliances
  as   $k_a=s_{11}+s_{12}$ for the cubic system and 
  $k_a=s_{11}+s_{12}+s_{13}$, 
  $k_b=s_{22}+s_{12}+s_{23}$, 
  $k_c=s_{33}+s_{13}+s_{23}$ for the orthorhombic system \cite{Paufler1986}.}
\label{bmeos}
\centering
    \begin{tabular}{llcc}
      \hline\noalign{\smallskip}
      & & exp.& calc.  \\ 
      \noalign{\smallskip}\hline\noalign{\smallskip}
      & b$_0$ [GPa]& 51.0(1.2) & 52.43(9) \\ 
      $\alpha$-PbS & b' & 4.3(9) & 4.69(2) \\
      & k$_a$ [GPa$^{-1}$]& 0.0063(8) & 0.00628  \\ 
      \noalign{\smallskip}\hline\noalign{\smallskip}
      & b$_0$ [GPa]& 30.9(4) & 25.6(6)\\
      & b' & 4(fixed) & 6.9(2)   \\ 
      $\beta$-PbS & k$_a$ [GPa$^{-1}$]& 0.0049(1) & 0.0069  \\
      & k$_b$ [GPa$^{-1}$]& 0.0032(4) & 0.0095  \\
      & k$_c$ [GPa$^{-1}$]& 0.0055(2) & 0.0024  \\
      \noalign{\smallskip}\hline
    \end{tabular}
\end{table}

\begin{figure}[!htb]
\centering
(a)  \includegraphics*[bb=20 28 440 442,height=0.75\columnwidth]{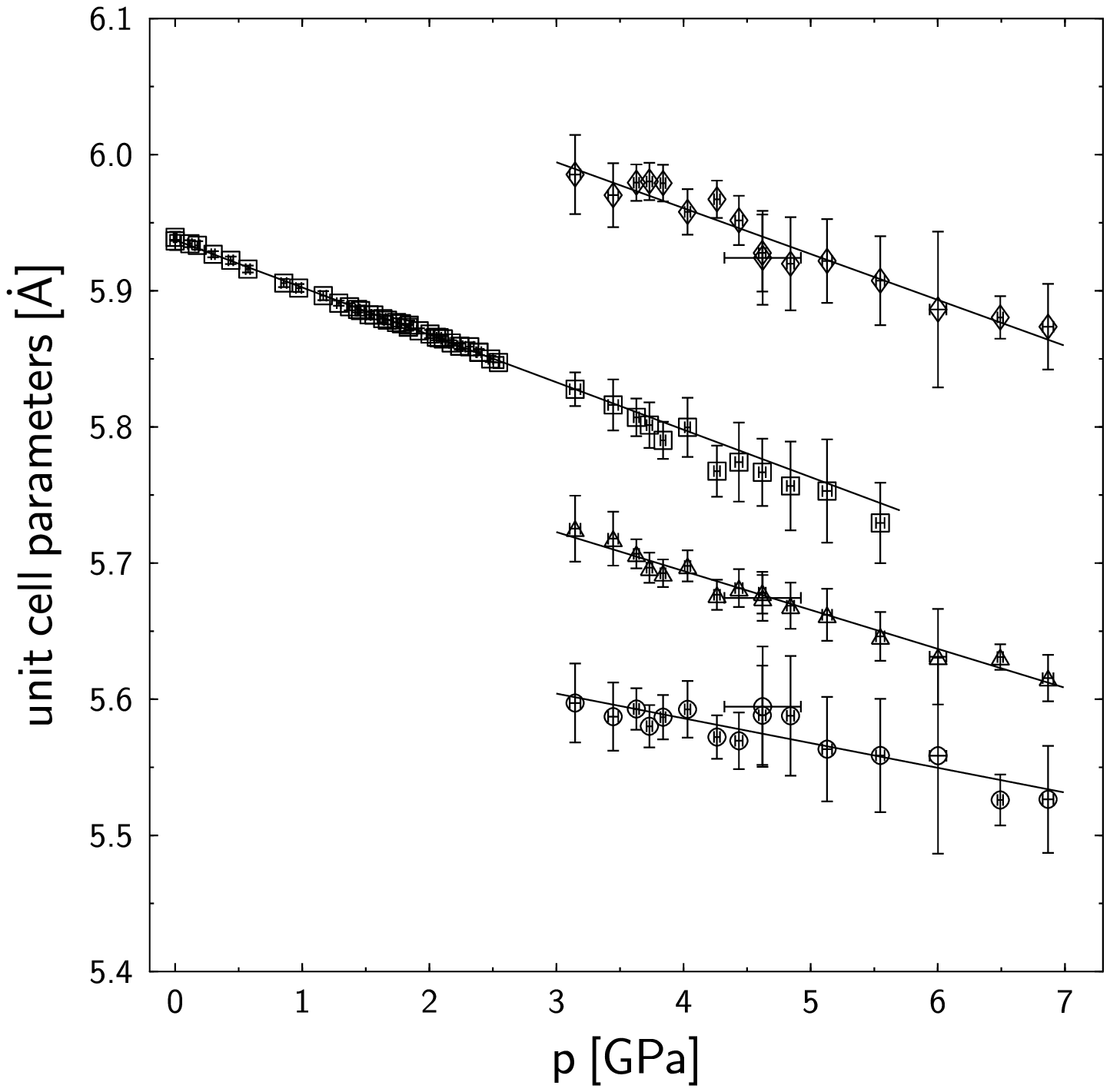}\\
(b)  \includegraphics*[bb=30 40 462 456,height=0.75\columnwidth]{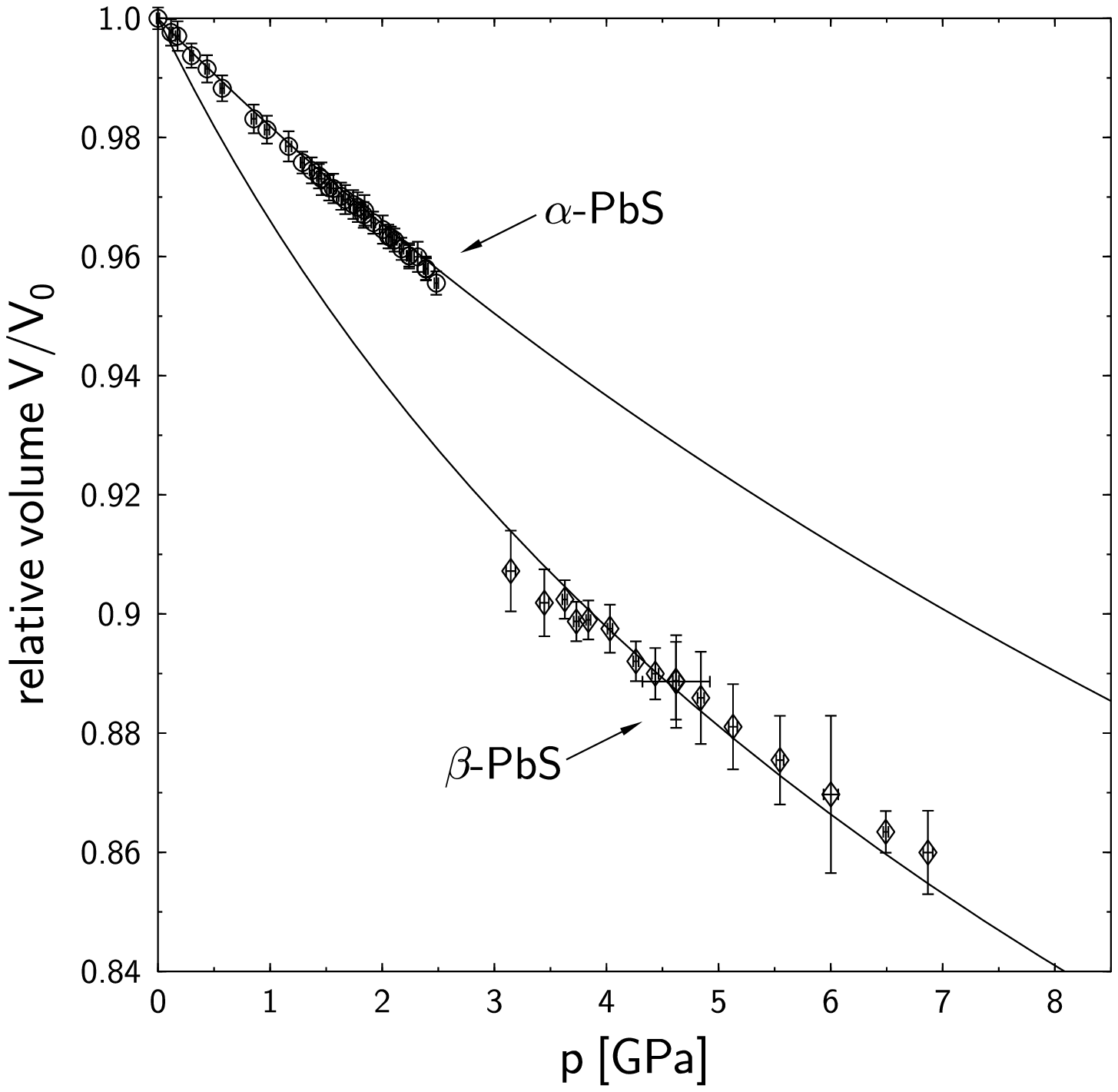}
  \caption{
    (a) Pressure dependence of the experimentally determined cell
    parameters for $\alpha$- ($\square: a_c $) and $\beta$-PbS (
    $\triangle: \sqrt{2}c_o $, $\circ: \sqrt{2}a_o$, $ \diamond:b_o/2$
    ).  (b) Pressure dependence of the normalised unit cell volume of $\alpha$-($\circ$) and
    $\beta$-PbS ($\diamond$).  The lines in (a) correspond to linear
    fits to the experimental data points and in (b) to fits of a
    3$^{rd}$-order Birch-Murnaghan equation-of-state to the {\it ab
      initio} data.  }
\label{fig:exp-cell}
\end{figure}

The cell parameters of the orthorhombic $\beta$-phase are related to
those of the cubic $\alpha$-phase by 
\begin{math}
  a_o=\sqrt{2}a_c/2, \quad b_o=2a_c , \quad
  c_o=\sqrt{2}a_c/2,
\end{math}
with subscripts $o$ and $c$ referring to the orthorhombic and the
conventional cubic lattices, respectively. Figure~\ref{fig:exp-cell}a
shows the dependence of (pseudo-) cubic lattice parameters on  pressure for both phases.

The normalised volume changes are given in figure~\ref{fig:exp-cell}b.
Because of the narrow pressure range observed in the experiment the
volume at zero pressure V$_0$ of $\beta$-PbS could not be obtained
unambiguously from the equation-of-state (eos) fit.  Instead, the
volume at zero pressure from the {\em ab initio} calculations was used to normalise
the experimental data of the $\beta$-phase.

The non-zero components of the calculated elastic stiffness tensor
$(c_{ij})$ (in Voigt's notation, units GPa) are
$c_{11}=141(3)$,
$c_{12}=34(2)$  and 
$c_{44}=19.97(5)$
 for $\alpha$-PbS at ambient pressure.
They agree well with the available experimental data ($c_{11}=149$,
$c_{12}=35$  and 
$c_{44}=29$ GPa) determined at 5~K
\cite{Padaki81}.
For $\beta$-PbS no experimental data are available, the computed
values at 4 GPa are
\begin{displaymath}
\left (
\begin{array}{rrrrrr}
 105(5)  &  11(6)   & 67(5)    &  &  &   \\
         &  89(8)   & 30(3)    &  &  &   \\
         &          &  103(7)  &  &  &   \\
         &          &          & 21(1)   &  &    \\
         &          &          &         & 71.02(6)&  \\
         &          &          &         &         & 12(2)     \\
\end{array}
\right ).
\end{displaymath}
Linear bulk moduli determined from elastic
constants are 70(2) GPa for $\alpha$-PbS and 53(3) GPa for
$\beta$-PbS. The value of the linear volume bulk modulus of the
$\alpha$-phase is higher than that of the high-pressure
$\beta$-phase. This reflects the different compression behaviour in
the two phases (section~\ref{sec:compr}).  The bulk moduli $b_0$ and
their pressure dependencies $b^{\prime}$ were determined for $\alpha$-
and $\beta$-PbS by fitting Birch Murnaghan equations-of-state
\cite{Birch78} to the normalised volume data. The fit results are
given in table \ref{bmeos} together with the linear compressibility
data $k$.

The bulk modulus $b_0$, as well as the pressure dependence of the bulk
modulus $b^{\prime}$, for $\alpha$-PbS show excellent agreement between
experimental and calculated values. Also, the linear compressibilities
derived from a linear fit to experimental data and from calculated
compliances agree excellently. The limited pressure range of the
experimental data prevented us from determining $b^{\prime}$ for
$\beta$-PbS. Hence, the bulk modulus was determined with a second
order Birch Murnaghan equation-of-state ($b^{\prime}:=4$). It is well
known that $b_0$ and $b^{\prime}$ are highly correlated variables in
eos-fitting \cite{Angel2000}.  Consequently, $b_0$ is higher in the
experiment, since $b^{\prime}$ is smaller comparing the numbers between
experiment and theory.  The resulting bulk modulus is smaller than
data reported recently \cite{Chattopadhyay86}. However, direct
comparison is difficult since the authors of \cite{Chattopadhyay86}
used a different equation of state.

\begin{figure}[!bh]
  \begin{center}
    \includegraphics[bb=25 15 730 530,width=0.95\columnwidth]{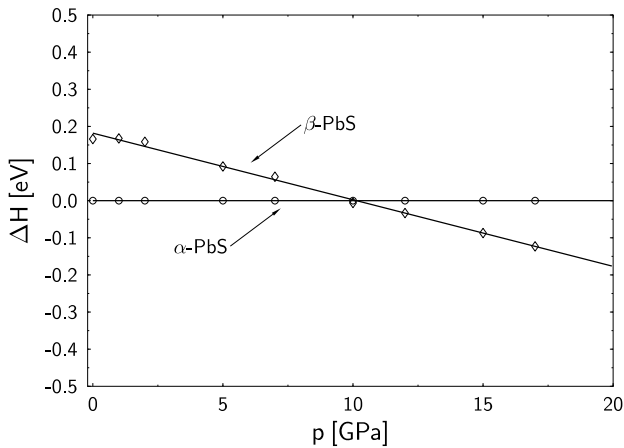}
    \caption{Pressure dependence of the molar enthalpy difference
      between $\alpha$- and $\beta$- PbS.}
    \label{fig:enthalpy}
  \end{center}
\end{figure}

\begin{figure*}[!t]
\centering
(a)  \includegraphics*[bb=10 30 280 330,height=6.5cm]{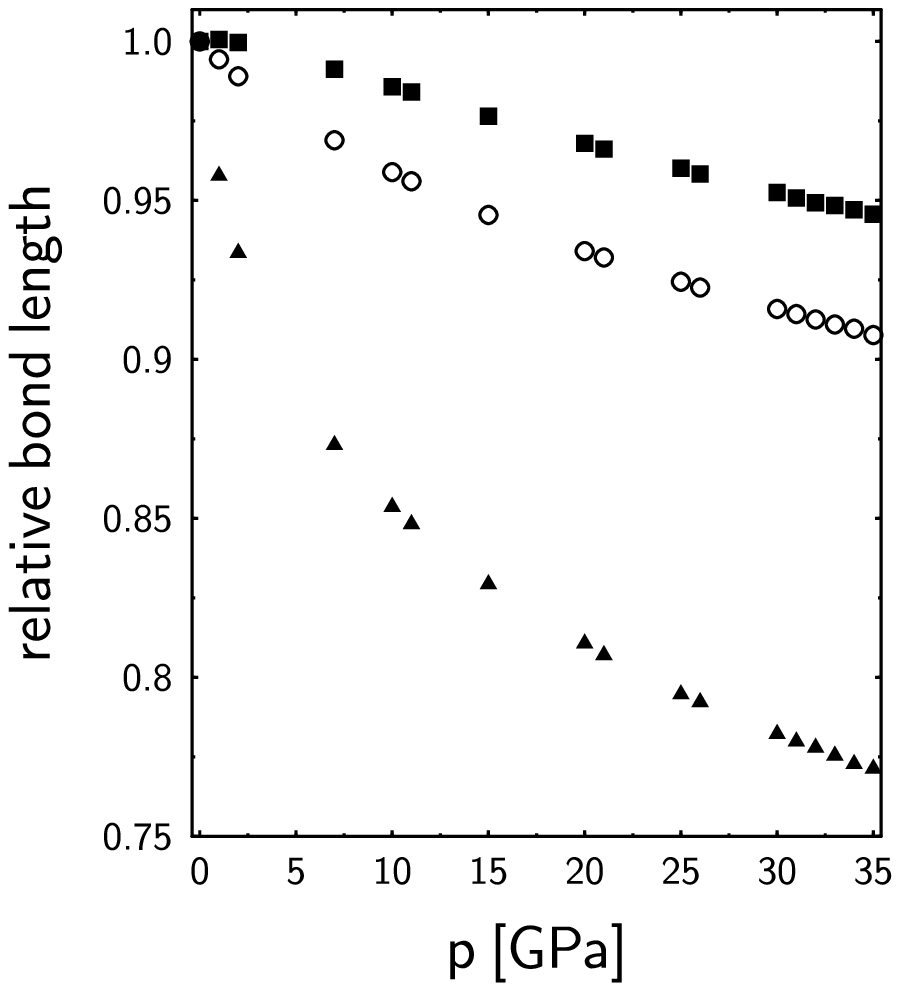}
(b)  \includegraphics*[height=6.5cm]{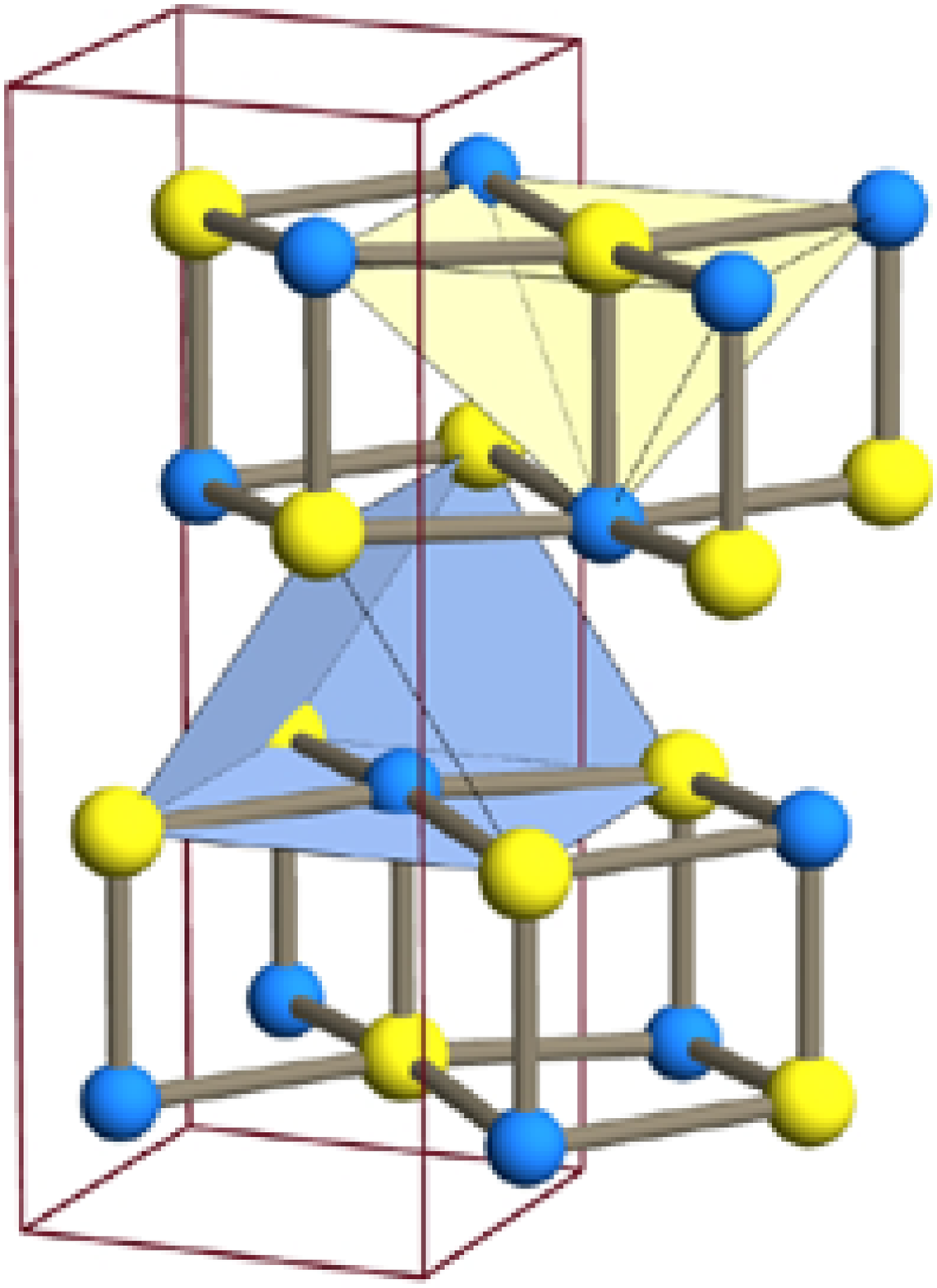}
(c)  \includegraphics*[height=6.5cm]{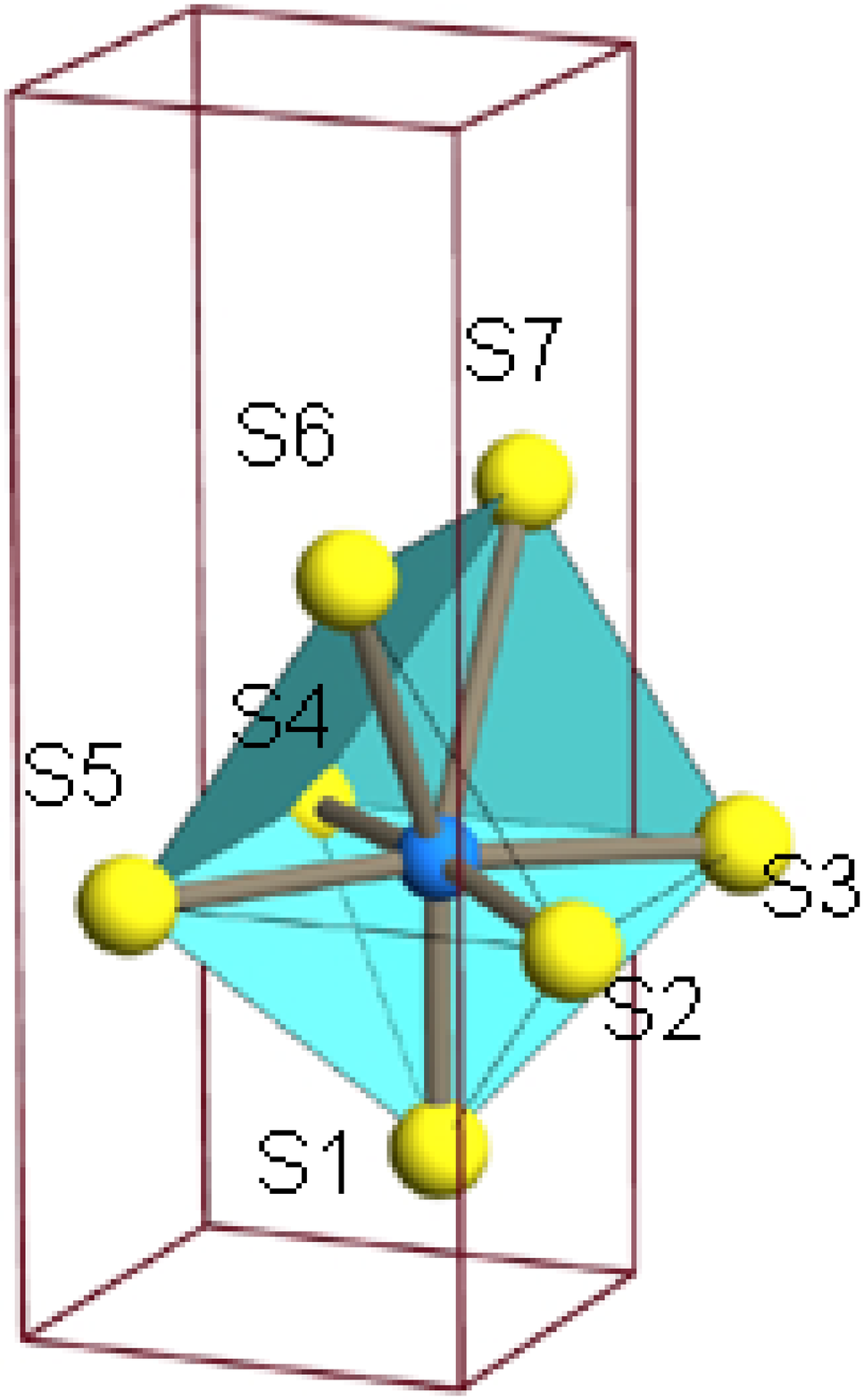}
\caption{(a) Variation of the Pb--S bond lengths in $\beta$-PbS from the
  {\em ab initio} calculations. In (b) and (c) coordination polyhedra
  and the crystal structure of $\beta$-PbS are drawn. Applying the
  labels given for the S atoms of the single-capped trigonal prism,
  shown in (c) the following bonds can be
  distinguished in graph (a): squares represent the
  Pb--S(1) bond pointing at the apex of the pyramid located within the
  bilayer in (b), circles correspond to Pb--S(2-5) bonds
  which constitute the base plane of the pyramid and one face of the
  prism, and triangles denote the Pb--S(6,7) bonds between
  adjacent bilayers and, being located inside the prism.}
\label{fig:calc-compression}
\end{figure*}

\subsection{Stability of the $\bm \beta$ phase}

As stated earlier, the quality of our diffraction data did not allow
the unambiguous assignment of the symmetry and, thus,  discrimination
between both possible structure types B16 and B33 for $\beta$-PbS.
By contrast, the quantum mechanical calculations made the distinction
possible, because only for the CrB-type structure (B33) convergence
was achieved. The enthalpy for this phase with respect to the
$\alpha$-phase is shown in figure~\ref{fig:enthalpy} as a function of
pressure.  A transition pressure of 9.6(1) GPa for the
$\alpha/\beta$-transformation follows from the crossover of the fitted
curves. This transition pressure is higher than the experimentally
determined values (onset of the transition at 2.6 GPa, this study; 2.2
GPa \cite{Samara62}). 
However, DFT-GGA calculations such as those presented here suffer
from systematic 'underbinding' and the neglect of temperature effects.
Furthermore, they refer to perfect ideal crystals. Hence, a discrepancy
of 7 GPa between observed and calculated transition pressures can be
tolerated.
Also, the main aim of such calculations is not to reproduce numbers,
but instead provide insight into the transition mechanism, and for
this purpose such calculations are well suited.


\subsection{Compression mechanism \label{sec:compr}}

In $\alpha$-PbS the lead atoms are sixfold coordinated by sulphur
atoms in a regular octahedral arrangement. The only structural
parameter in the B1 structure type not constrained by the space group
symmetry is the magnitude of the cell parameter. The Pb--S bond length
is $a_c/2$ and, of course, the compression mechanism results in
shortening of the bond length.

No major change in the linear compressibility between cubic and
orthorhombic PbS follows from the experimentally determined cell
parameters (Fig.~\ref{fig:exp-cell}b). However, Pb--S bond lengths
derived from the {\em ab initio} calculations
(Fig.~\ref{fig:calc-compression}a) clearly show that the compression
behaviour is an\-iso\-tro\-pic. Layered chalcogenides show a strongly
anisotropic compression with the strongest compression perpendicular
to the layers (e.g.  TiS$_2$ \cite{Allan1998}, SnS$_2$
\cite{Knorr2001-1}, NbS$_2$ \cite{Ehm2002-2}). The main compression
mechanism in these compounds is narrowing of the van-der-Waals gaps,
whereas the layers themselves are bonded predominately ionic.  The
structure of $\beta$-PbS can be thought of as a layer structure as well,
consisting of rock salt like bilayers, stacked along the $b$-axis of
the orthorhombic unit cell.  This is shown in
figure~\ref{fig:calc-compression}b where two simple coordination
polyhedra can be distinguished. Within the bilayers Pb is found in
a fivefold coordination for which a square pyramid is an idealised
geometry. The basal pinacoid face of the square pyramid forms together
with two sulphur atoms from the next bilayer an almost trigonal
prism. Figure \ref{fig:calc-compression}a shows the strongest
compression for this particular coordination polyhedron. The Pb--S(6,7)
bonds connect the layers.  The two different bonds within the bilayers
are less compressible. While the pyramids deform almost
isotropically,  the main compression mechanism is a
distortion of the trigonal prism, due to a change of the
Pb--S(6,7) bond lengths. 
This mechanism effectively influences the
length of the $b$-axis. Changes in the $a$ and $c$ axes are related to
lateral distortions within the bilayers, i.e. changes of
the Pb--S(2-5) bond lengths,  and of the angles between them.

Admittedly, a description of the coordination in $\beta$-PbS by two
different polyhedra is not fully appropriate since this approach
doesn't account for the increase of the coordination number with
pressure.  Alternatively, the coordination of the lead atoms can be
described as sevenfold with the geometry of a single-capped trigonal
prism \cite{Kepert1982}.  
The capped trigonal prism in
$\beta$-PbS is shown in Figure~\ref{fig:calc-compression}c.
The ideal geometry of a single-capped
trigonal prism requires that all apices are located on the surface of a sphere
and the edges $d_i$ fulfil the conditions given in table~\ref{geom}
\cite{Kepert1982}.
The deviation of this coordination polyhedron from ideal
can be quantified by applying e.g. the Pinsky and Avnir
continuous-symmetry-measure (CSM) \cite{Pinsky1998}. Here a
simplified CSM formulation was used. 

\begin{table}
  \caption{Geometric conditions for an ideal single-capped trigonal
    prism \cite{Kepert1982}. Five different edges with index $i$ have to
    be distinguished occurring with frequency $n_i$ and different edge
    lengths $d_i$. The radius of the circumsphere is $r$ and
    labels are given according to Figure~\ref{fig:calc-compression}c.}
  \label{geom}
  \centering
  \begin{tabular}{cccl}
    \hline\noalign{\smallskip}
    $i$ & $d_i$ & $n_i$ & label \\
    \noalign{\smallskip}\hline\noalign{\smallskip}
    1  & 1.277$r$ & 4 & S1--S2,3,4,5\\
    2  & 1.297$r$ & 2 & S2--S3, S4--S5\\
    3  & 1.233$r$ & 4 & S2--S6, S5--S6, S3--S7, S4--S7\\
    4  & 1.195$r$ & 1 & S6--S7\\
    5  & 1.478$r$ & 2 & S2--S5, S3--S4\\
    \noalign{\smallskip}\hline
  \end{tabular}
\end{table}

From the five different edges $d_i$ radii $r_i$ were determined according to
table~\ref{geom} and from the weighted average radius
\begin{equation}
\bar{r}_w = \sum_{i=1}^5 n_i r_i / \sum_{i=1}^5 n_i,
\end{equation}
the edge lengths of the corresponding ideal polyhedron $d_i^{\ast}$
could be calculated.
The average distortion measure ADM is then given by
\begin{equation}
\mbox{ADM} = 
 \frac{
   \sum_{i=1}^5 n_i 
      \left( d_i-d_i^{\ast}
      \right )^2
   }
   {
     \sum_{i=1}^5 \left ( n_i d_i^{\ast}\right )^2
   }
\times 100
\end{equation} 
and the mean edge length distortion is
\begin{equation}
\bar{\Delta}_d = 
  \sqrt{
     \sum_{i=1}^5 n_i \left(d_i-d^{\ast}_i\right)^2
     }  
   / 
  \sum_{i=1}^5 n_i.
\end{equation}
The ADM is normalised with respect to the edge lengths of the ideal
polyhedron in order to compensate for the effect of the
compression. Hence, it is a dimensionless number. The compression of
the coordination polyhedron is represented by the change of of the
circumsphere's radius $\bar{r}_w$ with pressure. The mean edge length
distortion $\bar{\Delta}_d$ is a measure of the variance of the
edges.  The resulting values for selected pressures are given in
Table~\ref{cms}.  At low pressures the conformation of the coordination
polyhedron deviates most strongly  from that of an ideal single-capped trigonal prism
\cite{Kepert1982} (ADM=0).  With increasing pressure ADM and
$\bar{\Delta}_d$ indicate increasing regularity of the coordination
polyhedron. Above approximately 10 GPa some residual
deformation remains virtually constant.
The position of the central lead atom does not coincide with the
conjoint centroid of the seven sulphur ligands (the difference
$\delta_y$ is along the orthorhombic $b$ direction and given in table
\ref{cms}).  With increasing pressure the position of Pb tends towards
that of the sulphur's centroid. A similar behaviour has recently
been reported for $\alpha$-PbF$_2$ \cite{Ehm2002-3}, where Pb is
elevenfold coordinated by F in a tri-capped trigonal prism.

\begin{table}
  \caption{Distortion measures of the single-capped trigonal prism in
    $\beta$-PbS at selected pressures, obtained from DFT calculations.}
  \label{cms}
  \centering
  \begin{tabular}{ccclc}
    \hline\noalign{\smallskip}
    p [GPa]& $\delta_y$ [\AA] &  $\bar{r}_w$ [\AA]&  ADM &  $\bar{\Delta}_d$ [\AA]\\
    \noalign{\smallskip}\hline\noalign{\smallskip}
    0  &  .3594 &   3.201  & .0289 & .0702\\
    2  &  .2149 &   3.117  & .016  & .0512\\
    7  &  .1005 &   3.019  & .0114 & .0416\\
    10 &  .0692 &   2.979  & .0111 & .0405\\
    15 &  .0355 &   2.928  & .0102 & .0382\\  
    21 &  .0095 &   2.879  & .0105 & .0381\\
    25 &  .0099 &   2.852  & .0099 & .0367\\
    \noalign{\smallskip}\hline
  \end{tabular}
\end{table}

\subsection{Characterisation of the $\bm \alpha /\bm \beta$ phase transition}

Coexistence of the two phases between 2.5 and 6 GPa was observed.
Since the peak profile remained sharp throughout the whole pressure
range covered by the experiment, pressure inhomogeneities can be
excluded as the origin of this phase coexistence. Instead, it reflects
the first order character of the transition
(e.g.~\cite{Chattopadhyay84,Chattopadhyay86}). This is further
supported by the observed discontinuity in the molar volume of 4.7~\%
at the phase transition.

\begin{figure}[!tb]
\centering
\includegraphics[bb=230 50 460 660,width=0.5\columnwidth]{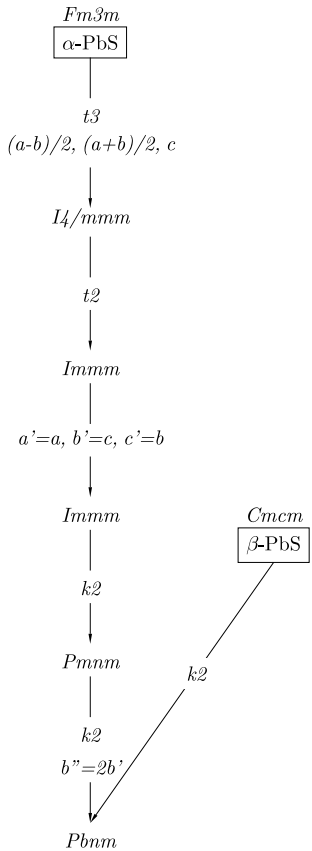}
\caption{Group/subgroup diagram for the $\alpha$/$\beta$ phase 
transition in PbS, showing a possible transition pathway from space group
$Fm\bar 3m$ to $Cmcm$ via the common subgroup $Pbnm$.}
\label{fig:tree}
\end{figure}

The transition is reconstructive. This is justified by i)
the change of the coordination number from six to seven, and ii.)
because there is no direct group/subgroup relationship between the two space
groups $Fm\bar 3m$ and $Cmcm$. Figure~\ref{fig:tree} shows the
symmetry relations in terms of {\it klassengleiche} (index $k$) and
{\it translationengleiche} (index $t$) subgroups together with the
changes of the unit cell dimensions. The crucial step here is the transition to space
group $Pbnm$ realised by a displacement $0 <\Delta a < 1/2$ of every
second (001)-bilayer along the orthorhombic $a$-, i.e. cubic $\langle
1\bar 10\rangle$, direction.  The space group $Cmcm$ is obtained as a
limit state, when $\Delta a$ takes the value of $1/2$.  It is interesting to note that the displacement is the
same as for the preferential glide system $\langle 1\bar
10\rangle\{001\}$ of NaCl. In the $\langle 1\bar 10\rangle$ directions
easy displacements of $\{001\}$ layers are possible because movement
of the ions occurs without having contact of equally charged ions.

The $\alpha/\beta$ transition in PbS is accompanied by a quadrupling
of the unit cell. The wave vector of the transition is $(0, 0, \pi/a)$
in the midpoint of the $\Delta$-line connecting the $\Gamma$-point at the zone
centre with the $X$-point on the boundary of the Brillouin zone of the
cubic face-centred lattice.

\section{Summary and conclusion}

The elastic behaviour of $\alpha$- and $\beta$-PbS was measured up to
6.8 GPa. Bulk moduli, elastic constants and linear compressibilities
were determined by DFT calculations and compared to experiment. The
symmetry of $\beta$-PbS was found to be $Cmcm$. The differences in the
compression behaviour between $\alpha$ and $\beta$-PbS can be
understood on the basis of the different crystal structures and
consequently, different compression mechanisms.  PbS bilayers as found
in the $\beta$-phase occur as  MX subsystem of the
chalcogenide misfit layer compounds  \cite{Rouxel95}. Structural
investigations under high pressure are not known for the misfit
layer compounds. Therefore, our
findings for the compression mechanism in $\beta$-PbS may serve at least as
model for the MX subsystem in misfit layer compounds. Of
course, for a better understanding of the elastic properties of the
MX-TX$_2$ multi-layer systems further investigations are required.

\section*{Acknowledgement}

This research was performed in the framework of the Kieler
Forschergruppe ''Wachstum und Grenzfl\"acheneigenschaften von Sulfid-
und Selenid-Schichtstrukturen'' funded by the German Science
Foundation DFG (De412/21-1).

\end{document}